\documentclass[useAMS]{mn2e}
\usepackage{psfig}

\def\lesssim{\mathrel{\hbox{\rlap{\hbox{\lower4pt\hbox{$\sim$}}}\hbox{$<$}}}}
\def\gtrsim{\mathrel{\hbox{\rlap{\hbox{\lower4pt\hbox{$\sim$}}}\hbox{$>$}}}}

\title[Jet opening angle and blazar parameters]{Influence of the jet opening angle on the derived kinematical parameters
of blazar jets having uniform and stratified bulk motion}
\author[Gopal-Krishna et al.] 
{Gopal-Krishna$^{1}$\thanks{e-mail: krishna@ncra.tif.res.in (G-K); samir@iucaa.ernet.in (SD); pronoys@iitk.ac.in (PS);  wiita@chara.gsu.edu (PJW)},  Samir Dhurde$^2$, Pronoy Sircar$^3$, Paul J.\ Wiita$^{4}$ \\  
$^{1}$ National Centre for Radio Astrophysics, TIFR,  Post Bag 3, Pune
University Campus, Pune 411
007, India\\
$^{2}$ Inter-University Centre for Astronomy and Astrophysics (IUCAA), Post Bag 4, Ganeshkhind, Pune 411 007, India\\
$^{3}$ Department of Physics, Indian Institute of Technology, Kanpur 208 016, India\\
$^{4}$ Department of Physics \& Astronomy, Georgia State University, P.O.\ Box 4106, Atlanta,
Georgia 30302-4106, USA}

\date{Accepted  2007 February 15;
      Received 2007 February 15;
      in original form  2006 July 22}

\pagerange{\pageref{446}--\pageref{452}}
\pubyear{2007}

\begin{document}

\maketitle

\label{firstpage}

\begin{abstract}
We present analytical modelling of conical relativistic jets, in order to
evaluate the role of the jet opening angle on certain key parameters that 
are inferred from VLBI radio observations of blazar nuclear jets. The 
key parameters evaluated are the orientation angle (i.e., the viewing angle)
of the jet and the apparent 
speed and Doppler factor of the radio knots on parsec scales. Quantitative
comparisons are made of the influence of the jet opening angle on the above
parameters of the radio knots, as would be estimated for two widely discussed 
variants of relativistic nuclear jets, namely, those having uniform bulk 
speed and those in which the bulk Lorentz factor of the flow decreases with 
distance from the jet axis (a `spine--sheath' flow). Our analysis shows that 
for both types of jet velocity distributions the expectation value of the jet
orientation angle at first falls dramatically with increases in the (central) 
jet Lorentz factor, but it levels off at a fraction of the opening angle for 
extremely relativistic jets.  We also find that the effective values of the 
apparent speeds and Doppler factors of the knots always decline substantially 
with increasing 
jet opening angle, but that this effect is strongest for ultra-relativistic
jets with uniform bulk speed. We suggest that the paucity of highly 
superluminal parsec-scale radio components in TeV blazars can be understood if their jets 
are highly relativistic and, being intrinsically weaker, somewhat less well 
collimated, in comparison to the jets in other blazars.

\end{abstract}

                                                                                   
\section{Introduction}
\label{sec:intro}

Although Very Long Baseline Interferometry (VLBI) monitoring of the radio knots in blazar jets has revealed
several sources containing knots with apparent speeds $v_{app}$ in excess of 25$c$ (e.g.,
Kellermann et al.\ 2004; Piner et al.\ 2006),
the typical speeds for blazars known to emit the highest energy $\gamma$-ray photons 
(TeV blazars) are found to be much more modest, with $v_{app} < 5c$, 
and their radio knots are often subluminal (e.g., Edwards \& Piner 2002;
Piner \& Edwards 2004). 
The glaring contrast with the very large bulk Lorentz factors in the 
parsec-scale blazar jets ($\Gamma > 30c$), as inferred from TeV flux 
variations (e.g., Krawczynski, Coppi  \& Aharonian 2002), has been highlighted by several 
authors (e.g., Giroletti et al.\ 2004; Gopal-Krishna, Dhurde \& Wiita 2004; Piner \& Edwards 2004, 2005;
Levinson 2006). Similar ultra-relativistic 
bulk Lorentz factors have also been inferred from the intraday radio 
variability of some blazars (e.g., 
Rickett, Kedziora-Chudczer \& Jauncey 2002; Macquart 
\& de Bruyn 2006) and, more directly, from VLBI measurements of the brightness 
temperatures of several blazar nuclei (e.g., Horiuchi et al.\ 2004; cf., Kovalev 
et al.\ 2005). The large mis-match between the estimates of $\Gamma$ 
derived from these different types of observations has led some authors 
to postulate a dramatic jet deceleration between sub-parsec and parsec 
scales (e.g., Georganopoulos \& Kazanas 2003; cf., Piner \& 
Edwards 2005).  An alternative approach has been to invoke a `spine--sheath'
configuration for the jets such that the fast spine close to the jet axis 
is surrounded by an appropriately slow moving sheath (e.g., Baan 1980; 
Komissarov 1990; Laing 1993; Meier 2003; Ghisellini, Tavecchio \& Chiaberge 2005).


These considerations recently led us to undertake an analytical study
which showed that the modest apparent speeds of the knots of blazars, 
which are mostly unresolved by VLBI, 
can be reconciled with the extremely relativistic bulk motion 
inferred for TeV, and some other, blazars as noted above, if one 
considers a modest full opening angle $(\omega \sim 5^{\circ}$--$10^{\circ})$ 
for the parsec-scale jets (Gopal-Krishna, Dhurde \& Wiita 2004, hereafter Paper I).  
We also showed that the actual viewing angles, $\theta$, of such 
{\it conical} jets from the line-of-sight can be substantially larger than 
those commonly inferred (e.g.\ Jorstad et al.\ 2005) by combining the flux variability and the VLBI proper 
motion data (Gopal-Krishna, Wiita \& Dhurde 2006, Paper II). Direct support 
for the assumption of {\it conical} parsec-scale jets comes from the VLBI 
imaging of the nuclear jets in the nearest two radio galaxies, M87 
(Biretta, Junor \& Livio 2002) and Centaurus A (Horiuchi et al.\ 2006).
 
Our focus in this paper is on making a quantitative comparison of the predictions for 
two widely discussed jet forms (i.e., those with uniform $\Gamma$ and 
those with velocities decreasing away from the jet axis, the spine--sheath 
types), of the extents to which some key parameters of the VLBI radio knots 
in blazar jets would be influenced by the jet opening angle.  The three 
main jet parameters examined are the apparent speed ($\beta_{app}$) of the 
knot, its Doppler factor ($\delta$) and the viewing angle ($\theta$) to the 
jet axis from our line-of-sight.
Although discussed specifically for blazar jets, the present results are 
also relevant for $\gamma$-ray burst sources (GRBs) which too are believed 
to arise from  extremely relativistic jets, with $\Gamma > 100$, and with even larger solid angles (e.g., 
M{\'e}sz{\'a}ros 2002; Liang et al.\ 2006, and references therein).

\section{Computation of the apparent parameters for conical jets}
\label{sec:uniform and spine--sheath jets}

Following the analytical prescription of Papers I \& II, we approximate a 
radio knot with a circular disc-shaped region of uniform intrinsic 
synchrotron emissivity, and thus uniform surface brightness, presumably arising from a shock in the jet (e.g., Marscher 
\& Gear 1985; Hughes, Aller \& Aller 1985). See Fig.\ 1. 

\begin{figure*}
\hspace*{0.2cm}\psfig{file=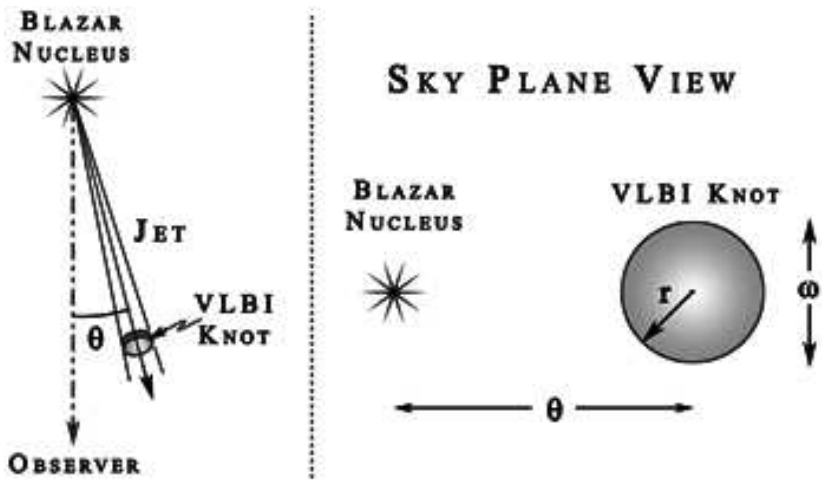}
\caption{Cartoon of a relativistic jet of mis-alignment angle $\theta$ and
full opening angle $\omega$.  A uniform radio knot in the jet appears as
a disc on the plane of the sky, the centre of which 
marks the location of the jet's axis and therefore lies at an angular 
separation $\theta$ from the line-of-sight to the nucleus.}
\end{figure*}

We define the angle between  the axis of the jet
and the line-of-sight from the stationary core of the blazar to the
observer to be
$\theta$ and the velocity parameter of the knot along the jet to be $\vec{\beta} \equiv \vec{v}/c$. Because we will consider jets with finite opening angles, and thus with a range of viewing angles to different portions of them, we use the more general form of the usual relations, 
\begin{equation}
\delta = \frac{1}{\Gamma (1-\vec{\beta} \cdot \hat{n})}, 
\end{equation} 
where  $\Gamma= (1-\beta^2)^{-1/2}$ and $\hat{n}$ is the unit vector along the sight line;
\begin{equation}
\vec{v}_{app} = \frac{\vec{v} \times \hat{n}}{(1-\vec{\beta} \cdot \hat{n})}
\end{equation}
In Papers I \& II we considered that each portion of the surface of the knot, denoted by the differential solid angle, $d\Omega'$, will have an angle
to the observer's line of sight that can range between $\theta - \omega/2$ and $\theta + \omega/2$.   Then the values
of $\delta$ and $v_{app}$ arising from each $d\Omega'$ patch vary (unlike for the commonly considered case where the jet is assumed to be a pencil beam with $\omega = 0$).  Here we generalize those results by considering the possibility that the magnitude
of the bulk velocity,
$c\beta$,  might also vary as a function of the angular distance ($r$) of that patch from the jet axis.  

The following basic relations 
are used for computing the {\it effective} values, $\beta_{\rm eff}$ and 
$\delta_{\rm eff}$, for a knot, as would be determined from VLBI observations which are unable to resolve the width of the jet (i.e., knot): 
\begin{equation}
S_{obs} = \int_{\Omega} \delta^{p}(\Omega')S_{em}(\Omega')d\Omega' \equiv
A(\theta)S_{em}; 
\end{equation}
\begin{equation}
\delta_{\rm eff} = A^{\frac{1}{p}}(\theta); 
\end{equation}
\begin{equation}
\vec{\beta}_{\rm app,eff} = \frac{1}{S_{obs}} \int_{\Omega} \vec{\beta}(\Omega^\prime)\delta^{p}
(\Omega^\prime)S_{em}(\Omega') d\Omega'.
\end{equation}  
Here, $S_{em}$ and $S_{obs}$ are the emitted and the (Doppler boosted) 
observed flux densities, respectively, $\Omega^{\prime} = \vec{\beta}\cdot{\hat{n}}/\beta$ denotes the location on the knot given by the direction cosine between the local velocity and the line of sight to the nucleus 
(which is no longer merely $\cos \theta$), $\Omega$ is the entire solid angle subtended by the  knot, $A(\theta)$ is the flux boosting factor averaged 
over the  radio knot's  cross-section, and $p$ is defined later in this section.   
Although we allow for a functional dependence for $S_{em}$ in Eqs.\ (3) and (5) our computations
assume that $S_{em}$ is actually
independent of $\Omega'$ (i.e., a uniform intrinsic surface brightness 
across the knot).  

In principle, several peaks in a VLBI jet could belong to a common
shock/knot yet appear separate only because of their different apparent
velocities (owing to their different velocity vectors within a cone).
In some cases, this approach has already suggested extremely large
bulk Lorentz factors ($> 100$, cf., Jorstad et al.\ 2004); however, there 
is usually a bias against accepting and reporting such extreme speeds. 
Moreover, such associations are unlikely to be the norm, since there is 
good evidence that individual bright VLBI knots arise from discrete 
ejection events in the nucleus, since their trajectories can mostly
be extrapolated to begin from the core at the times of emission
of discrete flares observed at millimetre wavelengths 
(e.g., Savolainen et al.\ 2002). Secondly, due to limited resolving power, 
the apparent motion of individual VLBI knots is commonly determined by 
monitoring the shift in the ``centroid" position of a given knot (e.g., 
Cohen et al.\ 2007).  To give expression to this 
practical situation, we have computed the flux weighted speed (Eq.\ 5). 
Clearly, this would no longer be required when the spatial resolution 
and the dynamic range of the VLBI image undergoes a substantial 
improvement.

The nominal viewing angle, $\theta$, to the jet axis 
measures the angular offset of the circular radio disc's centre (i.e., the jet axis)
from the 
direction of the AGN core (Fig.\ 1). The probability distribution of the viewing
angle, ${\cal P}(\theta)$ (see Eq.\ 7 below), is used to compute the expectation value of 
$\theta$ for any particular combination of $\Gamma, \omega, p$ and $q$, with 
the last two parameters defined below.
                                                                                
The integration was carried out numerically by dividing the circular radio
disc (i.e., the radio knot) of diameter $\omega$ and centred at  
$\theta$ into small pixels of 
angular size equal to $0.01 \omega$. For each pixel ($i,j$) in the disc, its 
angular offset from the AGN core was computed and combined with the jet's 
bulk Lorentz factor at that location (see below), in order to find: the Doppler factor $\delta_{i,j}$, the
apparent velocity, the vector
$\vec{\beta}_{app (i,j)}$; the flux boosting factor,  $A_{i,j}$, taken to be
$\delta_{i,j}^p$; and the product of the last two terms (which amounts to
the apparent velocity of the pixel, weighted by its apparent flux). 
Finally, for each of these parameters, the average taken over all the 
pixels across the radio disc was evaluated and this was taken taken 
to be the {\it effective} value of that parameter  (e.g., $A_{\rm eff}$) for the entire radio knot. Note that the 
effective value of $\delta$ for the knot is then $\delta_{\rm eff} = 
A_{\rm eff}^{1/p}$, where we have usually set $p =3$, widely used for compact 
discrete sources (e.g., Lind \& Blandford 1985) and we have assumed the knot 
to have a flat spectral index for simplicity; taking other values for the spectral index would make no qualitative difference to our results. We also consider the
case where $p = 2$, appropriate for smooth flows. The motion of each pixel is taken to be 
ballistic (see, e.g., Kaiser 2006).

\begin{figure*}
\hspace*{0.2cm}\psfig{file=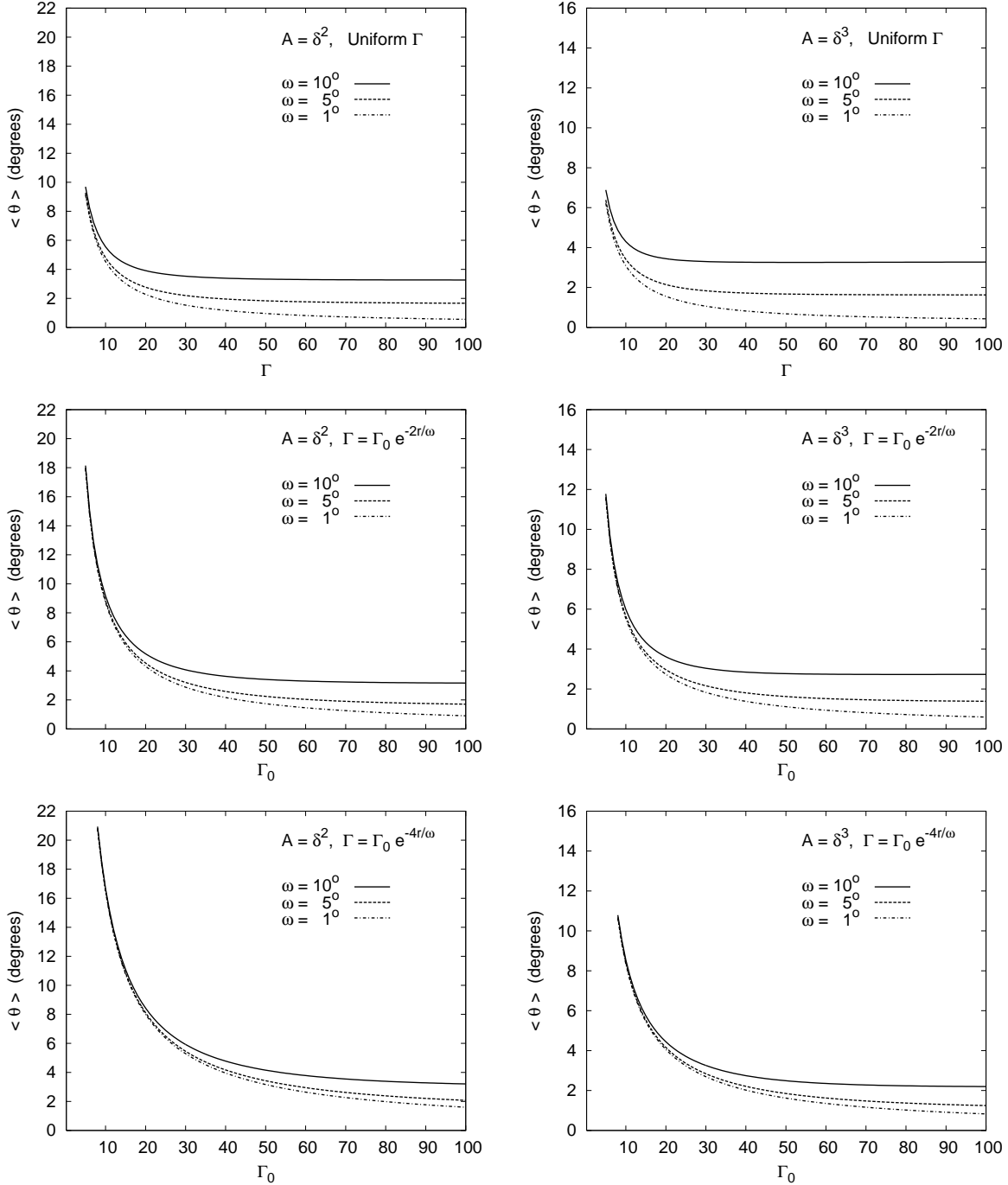}
\caption{Expectation values of the viewing angle, $\left<\theta\right>$, against the jet (central) Lorentz factor, $\Gamma_0$, for uniform (top)
and transversely structured (middle and lower panels) jets. The left panels 
correspond to $p=2$ and the right to $p=3$. Solid, dashed and dot-dashed lines are for 
$\omega = 10^\circ, 5^\circ$, and $1^\circ$, respectively.}
\end{figure*}
  
For the case of spine--sheath type jets, too, the (transverse) shock is assumed 
to cover the jet's entire cross-section with uniform intrinsic emissivity.  In the absence of any 
specific form argued for in the literature, we adopt an exponential approximation:
\begin{equation}
\Gamma (r) = \Gamma_0 e^{-2rq/\omega},
\end{equation}
where $r$ is the angular separation from the centre of the radio disc/knot
(i.e., from the jet axis).
We evaluated results for such stratified jets, taking two representative values for $q$ (1 and 2). The 
choice of the minimum $\Gamma_0$ allowed is then dictated by the constraint 
that the corresponding $\Gamma (r = \omega/2)$ remains above the physical 
limit of $\Gamma =1$.

Note that the
probability of finding a jet at a viewing angle $\theta$, in a
flux-limited sample, is given approximately by (Paper I; Cohen 1989):
\begin{equation}
{\cal P}(\theta)d\theta \propto \sin \theta~
A^{\frac{3}{2}}_{\rm eff}(\theta)~d\theta,
\end{equation}
where the exponent $\frac{3}{2}$ is the typical slope of integral source
counts at centimetre wavelengths (e.g., Fomalont et al.\ 1991)
and $A_{\rm eff}$ was taken from the present computations (see above). Eq.\ (7)
was numerically evaluated from $\theta = 0^\circ$ to $90^\circ$ and then used to
compute the expectation values $\left<\theta\right>$ for different combinations of
$\Gamma_0$, $\omega$, $p$ and $q$.

Results of the above numerical computations are displayed in Figs.\ 2--4, 
for both uniform $\Gamma$ jets ($q = 0$) and for the stratified jets ($q = 1$ and $2$). 
Fig.\ 2 shows the expectation values of $\theta$, computed for a range of 
$\Gamma$, for three representative values of $\omega ~(1^{\circ}, 5^{\circ}$ 
and $10^{\circ}$) and for $p=2$ and $p=3$. Figs.\ 3 and 4 display, for 
$p=3$ and $2$, respectively, the dependence of $\beta_{\rm eff}$ and 
$\delta_{\rm eff}$ on the jet opening angle $\omega$, for three values of $\Gamma_0 
~(20,~50$ and 100). Analytical expressions for the curves fitted to the computed
data points are given inside each of the panels. Note that any given point along 
these curves refers to the corresponding expectation value of $\theta$ for that 
point, as computed using the values of $\Gamma_0$, $\omega$,  $p$ and $q$ for
that point.   

\begin{figure*}
\hspace*{0.2cm}\psfig{file=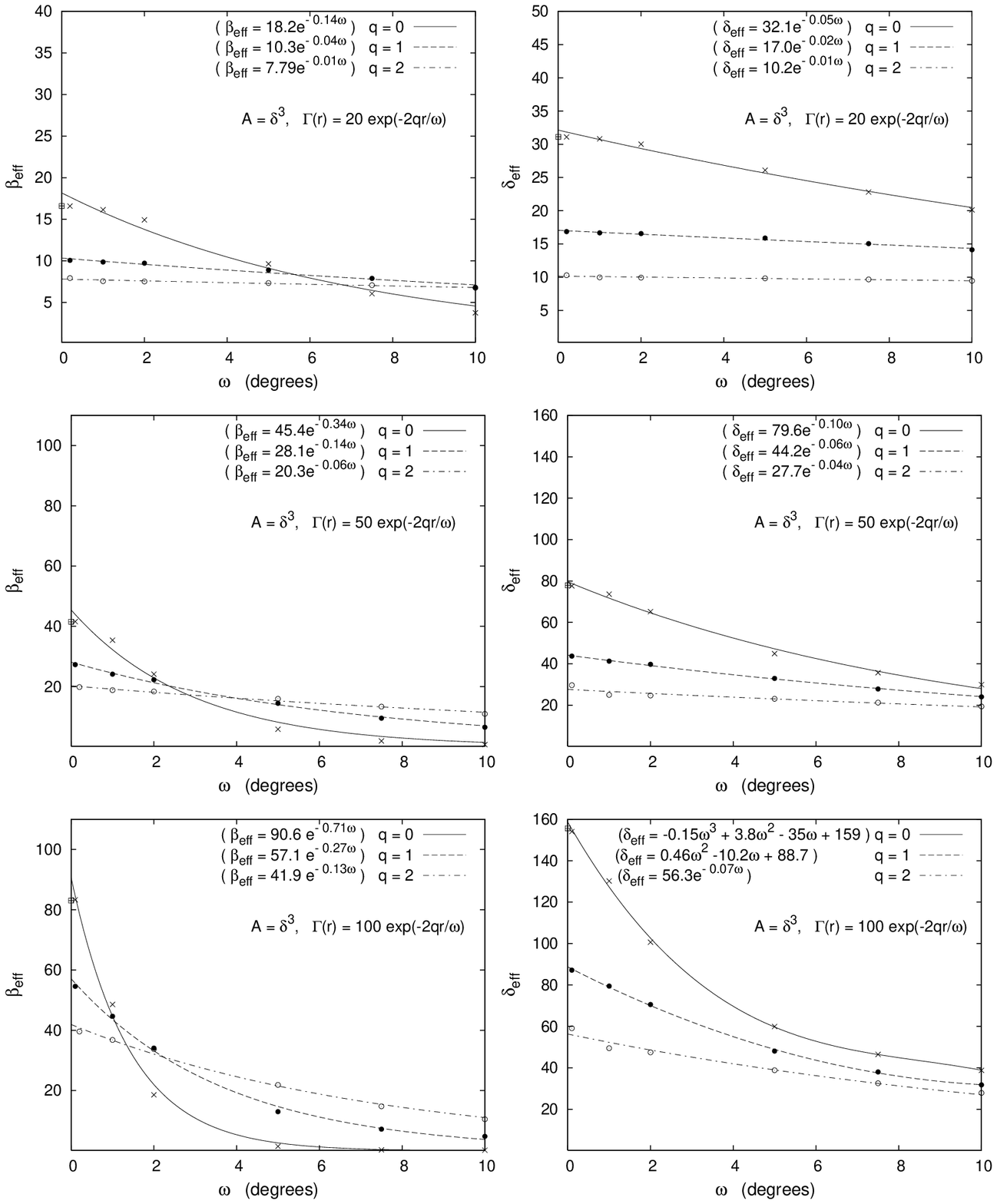}
\caption{The effective speeds (left) and effective Doppler factors (right) as functions of the full opening angle of the jet, with $p=3$.  The top, middle and bottom panels correspond to
$\Gamma_0 = 20, 50$, and $100$, respectively. The $\times$ symbols, closed circles and open circles are the computed values for $q=0,1,2$, respectively. The solid, dashed and dot-dashed curves give the corresponding labelled exponential function fits to the results except that polynomial fits are given to the $\delta_{\rm eff}$ values for the $\Gamma_0 =100$ and $q = 1,2$ cases. In each panel, the open square symbol gives the usual result for $\omega =0$ and constant $\Gamma$.}
\end{figure*}

\begin{figure*}
\hspace*{0.2cm}\psfig{file=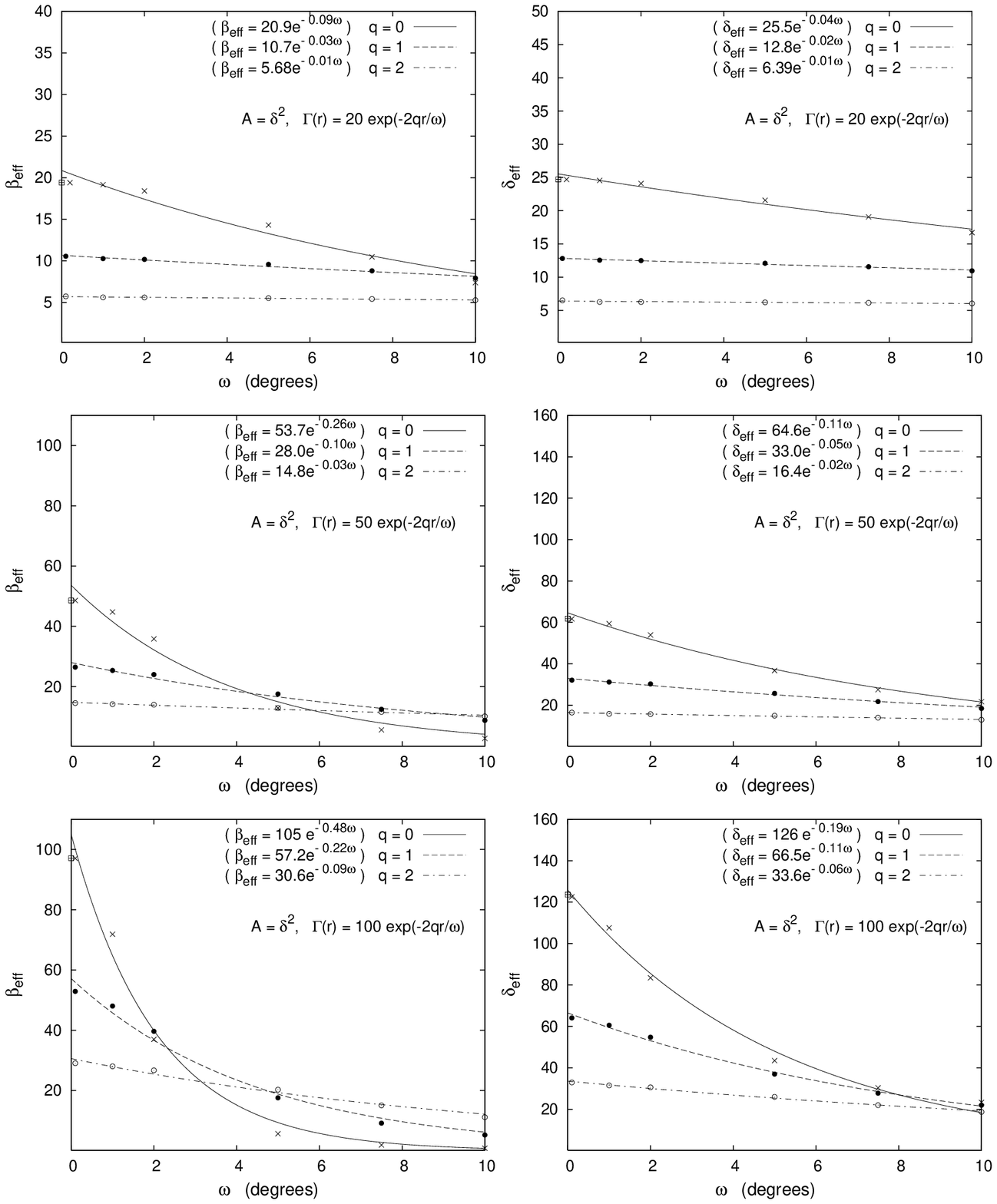}
\caption{As in Fig.\ 3 for $p=2$, except that exponential fits are made to all results.}
\end{figure*}



\section{Discussion}

Fig.\ 2 shows a steep initial decline of $\left< \theta \right>$ with 
increasing $\Gamma_0$, regardless of the chosen values 
of $\omega$ or $q$.  This decline continues until the regime of ultra-relativistic 
jets ($\Gamma_0 \gtrsim 30$) is reached, whereafter $\left<\theta\right>$ 
becomes essentially independent of $\Gamma_0$.  The near constancy of $\left< \theta \right>$ for extremely 
relativistic jets is particularly striking for the conical jets having a 
uniform $\Gamma$ and at least a moderate opening angle 
($\omega \sim 5^{\circ}$--$10^{\circ})$.  Further, in this 
ultra-relativistic regime, the tendency for $\left <\theta \right>$ to 
increase with $\omega$ is found to accelerate with the increase in the 
opening angle $\omega$. Interestingly, these trends found for the 
uniform $\Gamma$ jets are also shared by both representative forms of 
spine--sheath jets considered here.  Quantitatively, for $\omega = 5^{\circ}$, 
ultra-relativistic jets of all the three forms (i.e., $q = 0, 1$ and $2$), 
have $\left<\theta\right>$ in the narrow range from $1^{\circ}$ to $2^{\circ}$. The 
corresponding range of $\left< \theta \right>$ for $\omega = 10^{\circ}$ 
jets is $2^{\circ}$ to $4^{\circ}$, for all $\Gamma$s above $\sim 30$. From
Fig.\ 2, it is also evident that in a typical radio flux-limited sample of 
blazars, a larger opening angle of an ultra-relativistic jet would 
correspond to a considerably larger $\left< \theta \right>$; hence a milder 
fore-shortening due to projection is typically expected (see Paper II for 
an expanded discussion of this last point for uniform velocity jets).

We now examine the influence of $\omega$ on the {\it effective} apparent 
speed of a radio knot,  c$\beta_{\rm eff}$, and its {\it effective} 
Doppler factor, $\delta_{\rm eff}$ (Eqs. 4 \& 5).  These are displayed in Fig.\ 3 ($p=3$) and 
in Fig.\ 4 ($p = 2$) for the computed expectation values $\left< \theta \right>$ 
(which is a function of $\omega$, $\Gamma_0$, $p$ and $\delta$ and hence varies
along each curve, as explained above).  
Good analytical fits to simple exponential curves were found to all the
computed values of $\beta_{\rm eff}$ and they are quoted within those
figures.  Exponential functions also gave good fits to all but two of the 18 cases
considered for $\delta_{\rm eff}$; in those cases the displayed polynomial fits were 
found to be substantially better.

Firstly, it is seen that the decline of $\beta_{\rm eff}$ with $\omega$ 
is sharper for the knots associated with jets of higher $\Gamma$ (for both 
uniform and stratified types). On the other hand, for well collimated jets 
(i.e., $\omega \lesssim 0.5^{\circ}$), $\beta_{\rm eff}$ for the uniform $\Gamma$ 
case would typically be 1.5 to 2 times higher than for the $q=1$ and
between 2 and 4 times higher for the $q=2$ (the two spine--sheath 
cases).  Thus, the fastest {\it spine} component of the jet flow, which 
is near the jet axis, would be substantially concealed in the VLBI measurements. 

It is interesting to note that the much sharper fall of $\beta_{\rm eff}$ with 
$\omega$ found for the uniform $\Gamma$ case (especially in Fig.\ 2) ensures that 
already for modest jet opening angles, the value for $\beta_{\rm eff}$ of such 
jets would drop below the corresponding values for both of the spine--sheath 
models considered here. For $p = 3$ this crossover is seen to occur near $\omega 
= 6^{\circ}, 2.5^{\circ}$ and $1.3^{\circ}$,  for $\Gamma_0 = 
20, 50$ and $100$, respectively, such that beyond these modest jet
 opening angles, 
$\beta_{\rm eff}$ rapidly approaches mildly relativistic (or, even sub-relativistic) 
values in the case of extremely relativistic jets with uniform $\Gamma$. 
This steep fall of $\beta_{\rm eff}$ was first pointed out in Paper I 
for the case of uniform $\Gamma$ jets and we find 
here a similar, albeit milder, dependence for the stratified ultra-relativistic 
jets as well (Figs.\ 3 and 4). It is also worth noting that while for well
collimated ultra-relativistic jets $\beta_{\rm eff}$ is more strongly suppressed
(relative to the uniform $\Gamma$ case) when $\Gamma$ is more peaked towards the
axis (i.e., $q =2$); the opposite is found for the jets having a 
significant opening angle ($\omega \sim 2^{\circ}$ to $\sim 10^{\circ})$, with the exact 
cross-over value of $\omega$ depending 
on $\Gamma_0$ and $p$. In summary, provided the conical jet is moderately wide 
($\omega \sim 10^{\circ}$),  the measured apparent speeds of the VLBI 
knots would typically remain under $10c$, even if $\Gamma_0$ were extremely 
large ($\sim 100$), {\it not only for the uniform $\Gamma$ jets but 
even for the stratified jets}. 

The right columns of Figs.\ 3 and 4 illustrate the $\omega$-dependence of the 
`effective Doppler factor' ($\delta_{\rm eff}$) for the uniform $\Gamma$ 
jets and for the two stratified jet forms; recall that $\delta_{\rm eff}$ 
is the cube root (for the $p = 3$ case), or square root (for $p = 2$) of the 
flux boosting factor, $A_{\rm eff}$, averaged over the radio 
knot (Eq.\ 4). Again, each point of these profiles is meant for the corresponding 
value of $\left<\theta\right>$, as discussed above. It is seen that for cases of 
very good collimation ($\omega \le 0.5^{\circ}$), the uniform $\Gamma$ jets 
would typically have 2 to 4 times larger $\delta_{\rm eff}$ compared to the 
stratified jets, implying roughly an order-of-magnitude stronger Doppler boost. 
As expected for stratified jets, a sharper spine--sheath contrast (i.e., a larger
$q$) leads to 
a lower $\delta_{\rm eff}$. 

Further, as for $\beta_{\rm eff}$, a significant reduction 
in $\delta_{\rm eff}$ with $\omega$ is the typical expectation for both kinds of jet, 
the dependence being stronger for extremely relativistic jets, particularly 
the uniform $\Gamma$ type (Figs.\ 3 and 4). This leads to a situation where 
extremely relativistic jets of both uniform and stratified $\Gamma$ types 
end up with comparable $\delta_{\rm eff}$ values, once $\omega$ exceeds about 
$5^{\circ}$, although the $\delta_{\rm eff}$ values for the uniform jets 
do remain larger than those for the stratified jets over the wide parameter
space considered here.  These still very high values of $\delta_{\rm eff}$ mean that rapid
variability in TeV $\gamma$-ray emission is to be expected in any of our ultrarelativistic models as the
variability timescale is proportional to $\delta_{\rm eff}^{-1}$.  Significant changes
in TeV fluxes on sub-hour timescales have been reported for Mrk 421 (e.g., B{\l}a{\.z}ejoski et al.\ 2005). 

Often, VLBI observations reveal different apparent speeds for the radio knots 
in the same jet. This is usually interpreted by postulating that the knots reflect
`pattern speeds' which can be substantially different from the underlying speed 
of the jet (e.g., Vermeulen \& Cohen 1994; Cohen et al.\ 2007). 
In our picture, such variations can be readily understood in terms of 
surface brightness distributions across the different knots being dissimilar.
Moreover, the observed lack of sources with large apparent speeds but low
brightness temperatures (e.g., Kovalev et al.\ 2005; Homan et al.\ 2006) also 
suggests that the intrinsic pattern speed 
should be broadly correlated to the jet's bulk speed.

As noted in Paper I, for the case of fully collimated jets the bulk 
Lorentz factor would have to quickly drop by 1 to 2 orders of magnitude 
between sub-parsec (TeV) and parsec (radio--VLBI) scales in order to 
reconcile the very large Doppler factors inferred from the compactness 
argument with the marginally superluminal (even sub-luminal) motion 
observed for the VLBI knots (Sect.\ 1).  At present there is no 
direct observational evidence for such drastic deceleration (and the 
concomitant dissipation) occurring on sub-parsec scales in blazar nuclei. 
Rather, for  Cygnus A, where sub-pc radio knot motions can be measured from
millimetre VLBI and the jets are close to the plane of the sky (so that 
small changes in direction would not yield significant apparent speed 
changes) the evidence favors modest acceleration, not deceleration, on 
the sub-pc scale (Bach et al.\ 2005).  One aim of our model is to eliminate 
the need for such a massively rapid deceleration. 


A key question is: why is the Lorentz factor dichotomy so striking only 
for TeV blazars (e.g., Piner \& Edwards 2004)?  Essentially all blazars 
show two correlated peaks in their spectral energy distributions and are 
now usually classified by the frequency at which the lower frequency 
(synchrotron) peak is strongest (e.g., Padovani \& Giomni 1995; Sambruna, 
Maraschi \& Urry 1996).  Now, according to the popular scheme unifying  high 
energy peaked (HBL) and low energy peaked blazars (LBL) (e.g., Sambruna et al.\ 1996; Fossati et al.\ 1998), TeV emission, which is an HBL characteristic, would 
be more common among lower luminosity blazars.  This hypothesis has been supported by a
recent study, which indicates that the synchrotron peaks for powerful blazars (the Flat-Spectrum Radio Quasars) all remain below 
the rather high energy ($\gtrsim$ 1 keV) which is 
characteristic of HBLs and TeV blazars (Padovani 2007).

We also recall that 
extragalactic jets tend to be less well collimated for lower luminosity sources, although this is only well established 
on the kpc scales where their opening angles can be routinely
measured (e.g., Bridle 1984).  Opening angles on
sub-pc scales can  be measured fairly unambiguously only for very nearby sources such as 
M87 (Biretta et al.\ 2002) and Centaurus A (Horiuchi et al.\ 2006) where the jets lie rather close to the plane of the sky; 
both of these are indeed relatively weak sources with wide jets. 
On the scale of nuclear jets Blandford (1993) gives a theoretical argument
for this weak/wide correlation.  Assuming it does hold, 
then the correlation between HBL properties and intrinsically weaker jets would mean that the jet opening angle should 
be larger for TeV emitting jets. Likely consequences of this are: 
(a) the probability of detecting TeV blazars would be enhanced, since, 
even though the TeV emission itself is probably beamed very sharply, 
its effective 
beaming angle is in fact much larger, and actually more like the jet opening 
angle (see, Phinney 1985; Blandford 1993; Begelman, Rees \& Sikora 1994); 
(b) the wider jet would mean a bigger reduction in the apparent velocity of the 
radio knot (Paper I and Figs.\ 3 and 4).  We suggest that the combination of these
factors is probably responsible for the intriguing preference of TeV blazars
to display slower motions of their VLBI knots.

  
\section*{Acknowledgments}
We thank the referee, Amir Levinson, for suggestions which significantly improved
the presentation of these results.
PS would like to thank the Indian Academy of Science, Bangalore, for the
award of a summer student fellowship and the National Centre for Radio 
Astrophysics (NCRA-TIFR), Pune for the facilities provided for his summer 
project.  PJW is grateful for continuing
hospitality at the Princeton University Observatory; his efforts were
supported in part by NSF grant AST-0507529 to the University of Washington through a subcontract to Georgia State University.

                                                                                   \label{lastpage}
\end{document}